\documentclass[11pt]{article}

\usepackage{amsmath}
\usepackage{amssymb}
\usepackage{graphicx}
\usepackage{cite}
\usepackage{color} 

\date{}
\newcommand{\SI}{SI}
\newcommand{\mcd}{\rho} 
\newcommand{\scd}{\varrho} 
\newcommand{\bfr}{\mathbf{r}}
\newcommand{\fc}{\mathbf{c}}

\begin{document}

\begin{flushleft}
{\Large \textbf{Emerging Allee effect in tumor growth }}\\ \bigskip
Katrin B\"ottger$^{1}$, 
Haralambos Hatzikirou$^{2}$, 
Anja Voss-Boehme$^{1}$,
Miguel A. Herrero$^{3}$,
Andreas Deutsch$^{1,\ast}$
\\
\bf{1} Center for Information Services and High Performance Computing, Technische Universit{\"a}t Dresden, 01062 Dresden, Germany
\\
\bf{2} Center for Advancing Electronics, Technische Universit{\"a}t Dresden, 01062 Dresden, Germany
\\
\bf{3} Departamento de Matem\'{a}tica Aplicada, Facultad de Matem\'{a}ticas, Universidad Complutense, 28040 Madrid, Spain
\\
$\ast$ E-mail: andreas.deutsch@tu-dresden.de
\end{flushleft}

\section*{Abstract}
 Tumor cells develop different features to adapt to environmental conditions. A prominent example is the ability of tumor cells to switch between migratory and proliferative phenotypes, a phenomenon known as go-or-grow mechanism. It is however unclear how this particular phenotypic plasticity affects overall tumor growth. To address this problem, we formulate and study a mathematical model of spatio-temporal tumor dynamics where different responses to local cell density mediate the go-or-grow dichotomy. Our analysis reveals that two dynamic regimes can be distinguished. If cell motility is allowed to increase with local cell density, any tumor cell population will persist in time, irrespective of its initial size. On the contrary, if cell motility  is assumed to decrease with respect to local cell density, an Allee effect emerges, so that any tumor population below a certain  size threshold eventually extinguishes. These results suggest that strategies aimed at hindering  migration, for instance by enhancing contact inhibition, are worth to be explored as alternatives to  those mainly focused at checking tumor proliferation.

\section*{Author Summary}
 Controlling tumor growth remains a major medical challenge. Current clinical therapies focuses on strategies to reduce tumor cell proliferation. However, tumor cells may switch between proliferative and migratory behavior thereby allowing adaptation to environmental conditions such as local cell density. With a mathematical model, we determine the consequences of this migration-proliferation plasticity on tumor growth. Our work shows that small tumors can be driven to extinction by their intrinsic cell population dynamics if cell motility decreases with local cell density. In contrast, if cell motility increases with  cell density, the tumor inevitably grows. Our model suggests that regulation of cell migration plays a key role in tumor growth as a  whole, making this feature a potential target for clinical studies. 

\section*{Introduction}
 Tumor cells possess a remarkable phenotypic plasticity that allows adaptation to changing environmental conditions \cite{Klein2013,Meacham2013}. Well-known examples are the epithelial-mesenchymal transition \cite{Friedl2003,Friedl2011} and the shift from ATP generation through oxidative phosphorylation to an anaerobic, glycolytic metabolism, often referred to as  the Warburg effect \cite{Cairns2011}. A further example is phenotypic plasticity with respect to cell proliferation and migration \cite{Gao2005}, a phenomenon known as go-or-grow mechanism. Such a migration-proliferation dichotomy has been observed for non-transformed cells \cite{DeDonatis2008,Zheng2009} as well as in the course of tumor development \cite{Farin2006,Garay2013,Hoek2008,Jerby2012}. 
The precise molecular mechanisms underlying this dichotomy remain poorly understood. However, it has been suggested that the switch between migrating and proliferative phenotypes is dependent on the cells' microenvironment such as growth factor gradients \cite{DeDonatis2008}, properties of the extracellular matrix \cite{Giese2003} or altered energy availability \cite{Godlewski2010}. In this context, several mathematical models have shown that the migration-proliferation plasticity has a major impact on tumor spread \cite{Chauviere2010,Hatzikirou2012,Kim2011,Martinez2012,Pham2012,Tektonidis2011}. It turns out that local cell density, which is known to be correlated with the gradient of nutrients, secreted factors, oxygen or toxic metabolites \cite{Favaro2008,Vultur2004}, is a core factor for analyzing the dependence of the switch on tumor microenvironment.
However, while the consequences of density-dependent migration-proliferation plasticity on tumor spread have been explored already, potential effects of this plasticity type on tumor growth and persistence have not been investigated so far.

Here, we predict unexpected consequences of phenotypic plasticity between migratory and proliferative phenotypes for tumor growth with the help of mathematical modeling. Mathematical models have proven successful for analyzing various aspects of tumor dynamics, see for example \cite{Anderson2008,Byrne2010,Gatenby2003}. We develop a cellular automaton model which incorporates the microenvironmental effect by a local cell density dependence of the phenotypic switch. Model analysis reveals that two dynamically different regimes can be distinguished. If cell motility increases with local cell density, even a small initial tumor population will always grow. This regime can be associated to a biological situation where contact inhibition of cell migration (CIM) is downregulated. On the contrary, if cell motility decreases with local cell density, which is the case if CIM is present, tumor colonies which are small enough can be driven to extinction by the intrinsic cell population dynamics. We unveil that this behavior is a consequence of negative growth rates emerging at low densities, a phenomenon called Allee effect in ecology \cite{Courchamp1999}. Hence, the control of cell migration behavior has direct consequences not only for tumor dissemination, but also for intrinsic tumor growth, a fact that might open a window for new therapeutic approaches. In fact, our work predicts that tumors displaying this type of plasticity can potentially be driven to extinction if contact inhibition of migration is externally enhanced. However, loss of such type of inhibition will invariably lead to tumor persistence.

\section*{Materials and Methods}
\subsection*{Model definition}
We develop a stochastic, spatio-temporal cell-based model to study the effects of density-dependent phenotypic plasticity. In this way we account for single cell behavior that depends on the local, spatial microenvironment and for microscopic fluctuations which reflect cellular and microenvironmental heterogeneity. To do that, a discrete model, namely a lattice-gas cellular automaton (LGCA) is defined. LGCA models are well-suited to model cell-cell interaction and cell migration \cite{Deutsch2005,Dormann2002,Hatzikirou2008}. 

The LGCA algorithm is described on a discrete $d$-dimensional regular lattice $\mathcal{L}$ with periodic boundary conditions. Each lattice node $\bfr$ is connected to its $b$ nearest neighbors by unit vectors $\fc_i$, $i=1,...,b$, called velocity channels. The total number of channels per node is defined by $K \ge b$, where $K-b$ is an arbitrary number of channels with zero velocity, called rest channels. Each channel can be occupied by at most one cell at a time. We consider a tumor population of two mutually exclusive cell phenotypes, moving ($m$) and resting ($r$). Moving cells reside on the velocity channels, indexed by $i=1,...,b$, while resting cells are located within the rest channels, indexed by $i=b+1,...,K$, of the lattice. The total number of cells at time $k$ and node $\bfr$ is given by $n(\bfr,k)= n_m (\bfr,k) + n_r (\bfr,k)$, where $n_m$ and $n_r$ denote the moving and resting cell numbers, respectively. The parameter $K$ is a local cell number bound. This constraint is imposed since the maximal cell number in a given volume is limited in a biological tissue.

\begin{figure}[!ht]
	\begin{center}
		\includegraphics[width=0.7\textwidth]{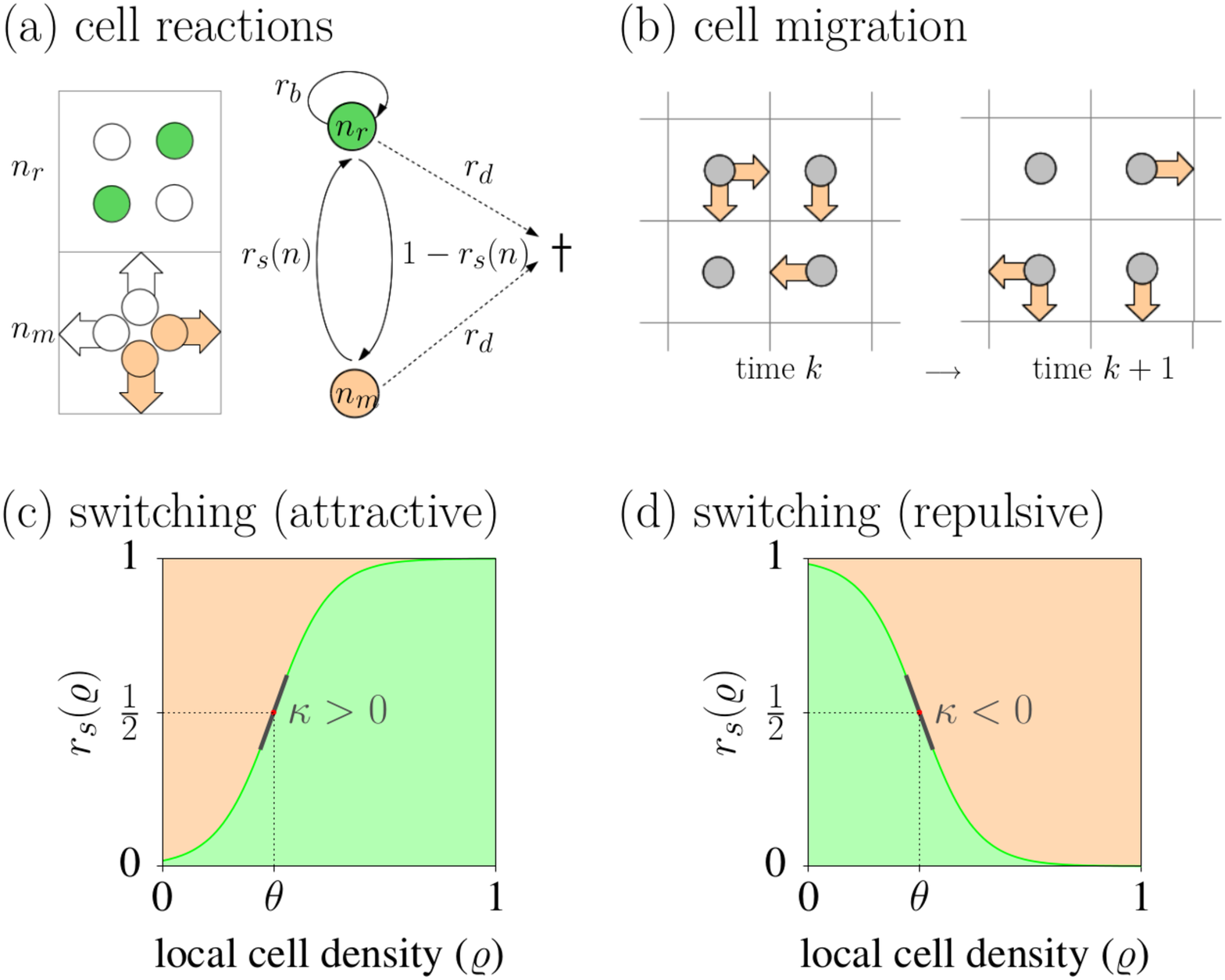}
	\end{center}
	\caption{
	CA model dynamics arise from repeated application of cell reactions and cell migration. (a) (left) Local state space at a given node is divided into rest channels and velocity channels. The cells in the rest channels, marked in green, are of proliferative phenotype. The cells in the velocity channels, marked in orange, have the migratory phenotype. White channels denote absence of tumor cells. (b) (right) Schematic illustration of model reactions. (b) Example of cell propagation in the LGCA model. Cells in the velocity channels before and after a propagation step; orange arrows denote the presence of a cell in the respective velocity channel. (c) and (d) Phenotypic plasticity in the CA model. Schematic illustration of the phenotypic switch probability $r_s$ which depends on the cell density $\varrho$ in the microenvironment. The sign of the phenotypic switch parameter $\kappa$ determines the dynamic regime, $\kappa >0$ results in attractive behavior while repulsive behavior arises for $\kappa <0$.
	}
	\label{fig-1}
\end{figure}

The time evolution of our model is defined by the following rules:
\begin{itemize}
	\item[(R1)] cells of both phenotypes undergo apoptosis with probability $r_d$,
	\item[(R2)] resting cells proliferate with probability $r_b$ unless all rest channels are occupied,
	\item[(R3)] cells change their phenotype with probability $r_s$ resp.~$1-r_s$ depending on the local node density,
	\item[(R4)] moving cells perform independent random walks.
\end{itemize}
In the LGCA, rules (R1)-(R4) are realized by applying three operators: A cell reactions operator changes the local cell numbers $n_r$, $n_m$ on each node according to (R1)-(R3). A reorientation operator randomly shuffles the configuration within the velocity channels at each node. By applying a propagation operator, moving cells are shifted one lattice unit in directions determined by their velocities. Both reorientation and propagation steps define cell movement, (R4). At each discrete time point $k$, the composition of the three operators is applied independently at every node on the lattice to compute the configuration at time $k+1$, see Fig.~\ref{fig-1}(a)-(b) and \SI.

We hypothesize that the phenotypic switch between proliferative and migratory cell behavior depends on the local cell density. We do not aim to reproduce the switch process in all intracellular detail. Rather we decide on simple cell-based mechanisms to provide a basic understanding of the underlying dynamics. In particular, we assume that the dependence on the cell density is monotonous. Then, two complementary types of plasticity can be distinguished: {\it attraction} towards or {\it repulsion} from highly populated areas. In the attraction case, cell motility decreases with local cell density, so that proliferation is favored in densely populated areas. In the repulsion case, cells tend to escape from highly populated regions, that is cell motility increases with local cell density, and proliferation is favored in sparsely populated areas. The switch probability $r_s(\varrho)$ (respectively $1-r_s(\varrho)$) that a moving cell becomes resting (or a resting cell becomes moving) is modeled as a sigmoidal shaped function $r_s:[0,1]\to (0,1)$ that depends on the cell density $\varrho =n/K$ at the given node and two parameters $\kappa \in \mathbf{R}$  and $\theta \in (0,1)$,
\begin{equation}\label{eq_switchFun}
	r_s(\varrho) = \frac{1}{2}(1+\tanh (\kappa (\varrho-\theta))), \quad \varrho \in [0,1].
\end{equation}
The absolute value of $\kappa$ specifies the intensity of the switch' density dependence while its sign determines whether the attraction case ($\kappa >0$) or the repulsion case ($\kappa<0$) is given. The parameter $\theta$ defines the critical cell density value at which the probabilities to switch from resting to moving and vice versa are equal. It determines the position of the switch, that is the density values with highest impact. We remark that the switch mechanisms are similar to those proposed in other studies on the density-dependent migration and proliferation dichotomy \cite{Chauviere2010,Pham2012}. 
A plot of the switching probabilities \eqref{eq_switchFun} is given in Fig.~\ref{fig-1}(c)-(d).

\subsection*{Model analysis}
We simulate the LGCA model on a two-dimensional lattice. We explore the effect of the switch intensity $\kappa$ and the switch position $\theta$ on the persistence of an invasive tumor population. To this end, we investigate the total population growth rates in the $(\kappa,\theta)$-parameter space and identify the parameter regimes for population survival and extinction. Proliferation and death probabilities are chosen such that $0<r_d\ll r_b\ll 1$. Simulations are performed on a square lattice with $10^4$ nodes. The initial model condition reflects a biological situation where the tumor is small and spatially constrained. At the initial time a fixed number of moving and resting cells per node is placed in a predefined radius from the lattice center. The initial cell density is varied by changing the percentage of occupied nodes within this radius.

\section*{Results}
\subsection*{Emergent population growth dynamics}
\begin{figure}[!ht]
	\begin{center}
		\includegraphics[width=0.5\textwidth]{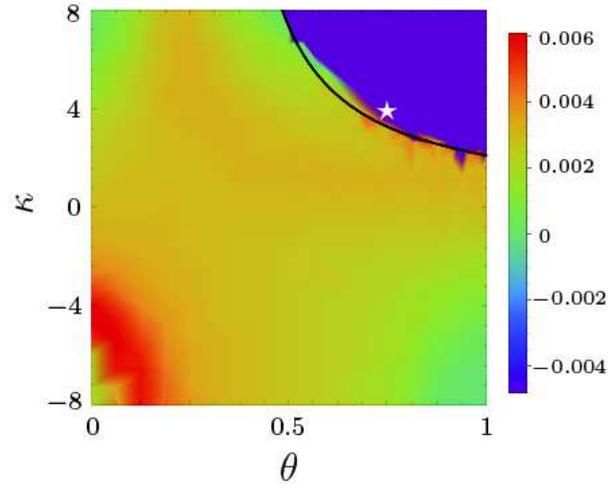}
	\end{center}
	\caption{
	Total population growth rate of the resting cell population depends on the phenotypic switch parameters $\kappa$ and $\theta$. Initially, in a radius of 10 [unit length] from the center one rest and one velocity channel per node are occupied. Parameters are $r_b = 0.2, r_d = 0.01$, $K = 8$ (1000 simulation time steps). The color indicates the total population growth rate of the resting cell population. The dark purple region denotes population extinction. The black curve is given by the inequality $r_s(\varrho) < r_d/r_b$, with $\varrho = 0.25$. The inequality derived in \eqref{eq_inEq} approximates the $(\theta,\kappa)$-parameter region for which population extinction is observed. The white star represents the specific $(\kappa,\theta)$-values for which subsequent analysis of the frequency of population extinction in Fig.~\ref{fig-3} is performed.
	}
	\label{fig-2}
\end{figure}

Figure \ref{fig-2} gives an overview of the observed cell population dynamics. In the repulsive case ($\kappa<0$), the population always persists, independently of the initial population density. In the attraction case ($\kappa> 0$), either survival or extinction may be observed, where the particular behavior is dependent on the specific values of $\kappa$ and $\theta$.

\begin{figure}[!ht]
	\begin{center}
		\includegraphics[width=0.7\textwidth]{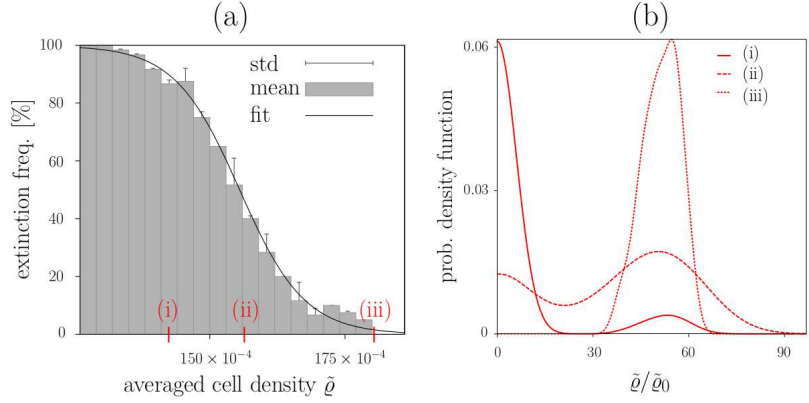}
	\end{center}
	\caption{
	Frequency of population extinction in the CA model depends on the initial population size. (a) Stochastic fluctuations lead to extinction or growth given a fixed small initial condition. The figure shows the frequency of extinction events depending on averaged cell density $\tilde\varrho = |\mathcal{L}|^{-1} \sum_{\mathbf{r}\in\mathcal{L}} \scd(\mathbf{r})$. (b) Probability density function of the averaged cell density after 5000 LGCA time steps for three different initial population sizes, derived by kernel density estimates (see \SI). For each initial configuration 40 simulation runs are performed. Model parameters are $\kappa = 4.4, \theta = 0.75, r_b = 0.2, r_d = 0.01, K=8$.
	}
	\label{fig-3}
\end{figure}

Further, we investigate the survival of low density populations in the transition region between both regimes ($\kappa \approx 0$). For a wide range of different initial population densities, we record the frequency of extinction events in each case. Sufficiently long simulation runs are performed to ensure that survival, when observed, is not a transient dynamical behavior. The results depicted in Fig.~\ref{fig-3}(a) show that the smaller the initial population density the higher the probability of population extinction. Above a critical initial population density, the population always survives. Additionally, we record the population size distributions after a number of time steps for different initial cell densities, see Fig.~\ref{fig-3}(b). One observes that low-density initial populations in the critical regime show bimodal stationary size distribution, indicating the possibility of either population extinction or persistence. 

The observed population behavior can be understood intuitively by considering the feedback mechanisms on cell behavior (proliferative or migratory) for the different types of phenotypic plasticity (attractive or repulsive). In the repulsion case ($\kappa<0$), increasing local cell density has a negative feedback on proliferation. In a sparsely populated environment, cells are predominantly resting. Cell replication leads to an increase in local cell density which in turn triggers the switch to a migratory phenotype. Migration of cells decreases local cell density which again triggers the switch towards the resting cell phenotype. As a consequence, proliferation and migration phases alternate, and the population always persists. In the attraction case ($\kappa>0$), increasing local cell density has a positive feedback on proliferation. Accordingly, cell proliferation leads to increased cell density which implies further proliferation. On the other hand, migration of cells locally decreases cell density that leads to more migratory cells. If the portion of resting cells in sparse environment is large enough, cell replication domains cell death. Thus, the positive feedback on proliferation might result in population growth. However, if cells in sparse environment almost exclusively migrate, they eventually die by apoptosis. Thus, the result is population {\it extinction}.

\subsection*{Phenotypic plasticity determines the existence of an extinction threshold}
The intuitive picture just described is supported by a mean-field theory. To this end, we derive a mean-field description of the cell-based model which was set up above. It allows us to analytically investigate the existence of an extinction threshold. The LGCA model is composed of a birth-death process describing the single-node cell reactions and a cell-movement process which describes the exchange of cells between neighboring nodes. Therefore, we derive mean-field approximations for each process separately first and then combine the resulting descriptions into a partial differential equation for the whole tumor growth process. We demonstrate that these descriptions allow to explain the behaviors observed in the simulation study and that they give additionally insight into the mechanisms of tumor persistence. 

If cell migration is neglected in the LGCA dynamics, the deterministic net changes occurring in the cell density of a given node between two consecutive times $k$ and $k+1$ are given by cell reactions only. Scaling time and transition rates appropriately such that the microscopic time $k$ corresponds to the macroscopic time $t=\tau k$, $\tau \ll 1$, one obtains two ordinary differential equations for the migratory and proliferative cell density, respectively. However, such system is hard to analyze analytically because of non-linearities which arise from the phenotypic switching. In order to facilitate analytical treatment, like bifurcation analysis, we assume that the switch dynamics is much faster than cell proliferation and death. In that case, we consider the system to be in equilibrium with respect to the switching. Hence, for low cell density, the fractions of cells that are in the resting and moving compartments are given by $r_s(\rho)$ and $1-r_s(\rho)$, respectively (see \SI). The overall macroscopic growth term of the LGCA model can then be approximated by 
\begin{equation}\label{eq_singleGrowth}
        F(\mcd)= R_b r_s(\mcd) \mcd \left( 1-\mcd \right) -R_d \mcd,
\end{equation}
with $\mcd:= \mcd_m+\mcd_r$, where $\mcd_m$ and $\mcd_r$ is the mean cell density of moving and resting cells, respectively, at a given position. The parameters $R_b$ and $R_d$ relate the models' proliferation and death parameter, $r_b$ and $r_d$, to the corresponding real time step length $\tau$. If the average cell cycle time of a cell is given by $T_b$ and the average life time of a cell by $T_d$, then $R_b=1/T_b\approx r_b/\tau$ and $R_d=1/T_d\approx r_d/\tau$ (see \SI).

\begin{figure}[!ht]
	\begin{center}
		\includegraphics[width=0.5\textwidth]{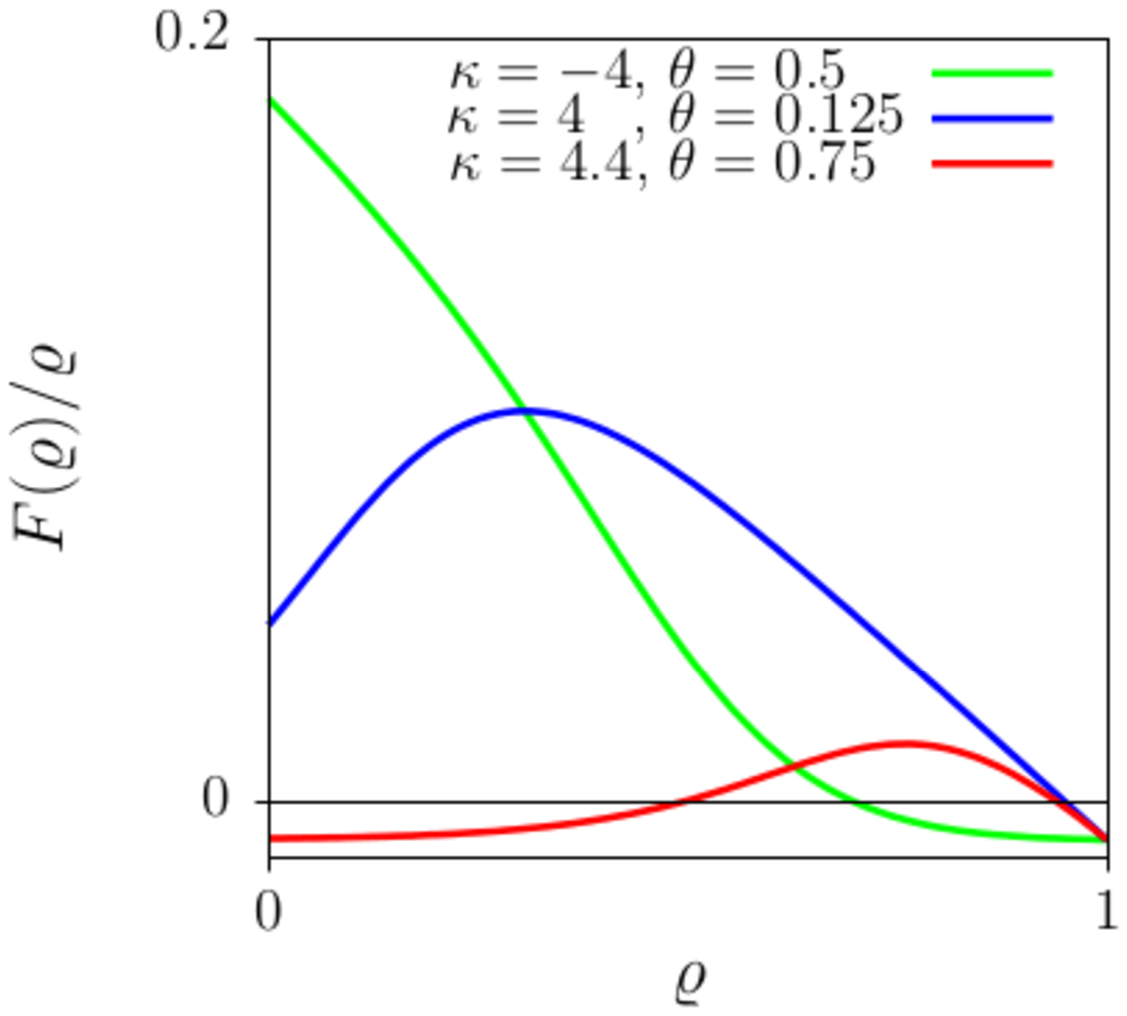}
	\end{center}
	\caption{
	Per capita growth rate of the mean-field description \eqref{eq_singleGrowth} depends on the type of phenotypic plasticity. In the repulsion case ($\kappa<0$), the per-capita growth rate is always positive (green line). In the attraction case ($\kappa>0$), the per-capita growth rate is reduced at low density  (blue line) and can even become negative (Allee effect, red line). Model parameters are $r_b = 0.2, r_d = 0.01, K=8$.
	}
	\label{fig-4}
\end{figure}

Stability analysis of the macroscopic net growth term $F(\mcd)$ shows that the behavior depends mainly on the type of phenotypic plasticity, attractive ($\kappa >0$) or repulsive ($\kappa<0$) (see \SI). More precisely, one finds that there are essentially two regimes, a monostable one for $\kappa <0$ and those values of $\kappa>0$ for which $r_s(\scd)r_b > r_d$ for small density values $\scd$, and a bistable one for $\kappa >0$ and 
\begin{equation}
	r_s(\scd)r_b < r_d. \label{eq_inEq}
\end{equation}
In the monostable regime, the cell density stabilizes at high density values where the exact location of the stable state is determined by the logistic growth restriction. In the bistable regime, the extinction state is stable, additionally to the stable high-density state that is due to the logistic growth restriction. The critical region where $\kappa >0$ and $r_s(\scd)r_b = r_d$ for small density values $\scd$ is depicted in Fig.~\ref{fig-2} for $\scd=0.25$. It shows good agreement with the simulation results. The stability of the extinction state in the bistable regime is due to the fact that the per capita growth rate $F(\mcd)/\mcd$ is negative for small density values, see Fig.~\ref{fig-4}. Such negative density-dependence is termed Allee effect in the ecology literature and has been attributed strong impact on population persistence and invasion properties \cite{Taylor2005}. Here, the Allee effect emerges as a consequence of the phenotypic plasticity with respect to migratory and proliferative tumor phenotypes.

\subsection*{Phenotypic plasticity leads to density-dependent migration}
The mean-field approximation for the LGCA cell movement process, detailed in \SI, is also derived under the assumption that the switch dynamics are much faster than cell proliferation and death. Since then, for low cell densities, a density-dependent portion $r_s(\rho)$ of cells is in the resting state, the diffusion coefficient turns out to be density-dependent. The LGCA migration process is isotropic with respect to the principal lattice direction. Along any direction, the diffusion equation for the mean-field cell migration in the macroscopic limit is given by 
\begin{equation}\label{eq_Diff}
	\partial _t \mcd = \partial_{x}\left(D(\mcd) \partial_x \mcd\right),\quad t\ge 0.
\end{equation}
where the diffusion coefficient satisfies 
\begin{equation} \label{eqn:DiffCoeff}
	D(\mcd) = D\left( \frac{1- r_s(0)}{2} - r_s'(0)\mcd - \frac{3}{2}r_s''(0)\mcd^2 \right),\quad \rho \ll 1,
\end{equation}%
with $D$ being the diffusive scaling constant that is related to the single cell motility.

Combining the mean-field descriptions for the cell reactions and the cell migration processes, we obtain a single partial differential equation (PDE),
\begin{equation}\label{eq_PDE}
	\partial _t \mcd = \partial_{x}\left(D(\mcd) \partial_x \mcd\right) + F(\mcd).
\end{equation}
where the macroscopic growth term is given in \eqref{eq_singleGrowth} and the density-dependent diffusion coefficient is given in \eqref{eqn:DiffCoeff}.

Analysis of \eqref{eq_PDE} with respect to the existence and speed of traveling wave solutions allows to characterize the tumor cell population dispersal that results from the phenotypic plasticity between proliferative and migratory phenotypes. The findings depend most notably on the stability behavior of the reaction term which in turn is a result of the presence (bistable reaction term) or absence (monostable reaction term) of the Allee effect. The dynamics of a simplified, semilinear version of \eqref{eq_PDE} where the diffusion coefficient $D(\mcd)$ is constant has been studied both for monostable and the bistable reaction terms \cite{Murray2002}. In particular, it was shown that such a semilinear equation admits front travelling waves of the form $u(x,t)=U(x-ct)=U(z)$ with $U(z)=1$ as $z\rightarrow -\infty$ and $U(z)=0$ as $z\rightarrow +\infty$ for some wave speed $c$. In the bistable case, the wave speed is uniquely determined and can be positive or negative, depending on the precise form of $F(\mcd)$. In contrast, in the monostable case, infinitely many (only positive) wave speeds are possible, all of which should satisfy an explicit lower bound \cite{Mikhailov1994,Murray2003}. A related fact is that in the monostable case invasion cannot be stopped once it starts \cite{Aronson1978,Hamel2012}, whereas sufficiently small initial populations may eventually become extinct in the bistable case \cite{Du2010}. The full equation \eqref{eq_PDE} where the diffusion coefficient depends on the cell density cannot be analytically treated yet.

\section*{Discussion}
To study the effect of plasticity between migratory and proliferative behaviors on tumor growth, we developed a cellular automaton model. The trigger for the phenotypic switch was assumed to depend on the microenvironment, in particular the local cell density. We found that this migration-proliferation plasticity has dramatic consequences for tumor growth. In more detail, two parameter regions with respect to the migratory cell behavior can be distinguished where fundamentally different tumor growth dynamics at the tissue scale are observed. In one case, called repulsive regime here, the tumor cell population will inevitably grow. In the other case, called attractive regime, we identified parameter regions, where sufficiently small tumors die out and tumor growth is only observed if the tumor size is above a certain threshold. We revealed that the extinction behavior is a consequence of the negative net cell growth rate at low cell densities, a phenomenon known as Allee effect in ecology. 

The Allee effect emerges from the specific regulation of the migration-proliferation plasticity at the cellular scale and has not been assumed {\it a priori}. It has been overlooked so far in the context of tumor growth and persistence. Since the Allee effect has been shown in ecology to change optimal control decisions, costs of control and the estimation of the risk posed by potentially invasive species \cite{Taylor2005}, we believe it to be critical for tumor growth control as well. The Allee effect observed here is actually stochastic, displayed by the discrete cellular automaton model. This means that it is not an artifact arising from the mean-field approximation where stochastic effects are averaged out. It reflects stochastic fluctuations which may be dominant for populations not too large, as it is the case for an early-stage tumor.

It can be expected from ecological studies that the Allee effect has also implications for tumor spread.  Here, we derived a mean-field description for the tumor cell density which is given by a reaction-diffusion equation with density-dependent diffusion coefficient and a potentially bistable reaction term. It will serve as a basis to investigate the resulting spreading behavior. Existing theoretical results, for example in \cite{Sanchez2010}, indicate that equation \eqref{eq_PDE} has a very rich behavior. In contrast to related reaction-diffusion systems with constant diffusion coefficient, which would apply to a non-density-dependent switch, standing waves and oscillatory front solutions can be expected for equation \eqref{eq_PDE} besides monotonic traveling wave front solutions. The numerical study of a model similar to equation \eqref{eq_PDE} with bistable growth term shows that the inclusion of dispersal effects can lead to different propagation speeds dependent on the initial cell densities in the simulations \cite{Cherubini2012}. Thus, we speculate that the Allee effect significantly affects tumor dissemination.
 
The microenvironmental influence on the phenotypic switch between moving and proliferative cell behaviors is incorporated in our model through a local cell density dependence. This is a plausible assumption since further potential environmental influences such as nutrient and oxygen supply, molecular signal gradients or other cell-cell interactions are mediated through and correlate with the local cell density. Therefore, we expect that our model reveals key features of population dynamics when a migration-proliferation plasticity exists. Note that this reduction of the underlying complexity is not a drawback of the model but allows to reveal inherent organizational principles. 

The stochasticity of our model incorporates heterogeneity of the microenvironment in its simplest form. Studying the implications of heterogeneity was not the focus of our investigation. However, ecological studies show that environmental heterogeneity has minor effects in growth dynamics where the Allee effect is present \cite{Dewhirst2009,Vergni2012}. Further investigations are required to analyze the importance of heterogeneity for specific tumors. 

Our results might have implications for the interpretation of recent experiments on tumor progression. It has been observed for low-grade cell line cultures, {\it in vivo} and {\it in vitro} that they have low chances of persistence and low reproducibility \cite{Huszthy2012,Tilkorn2011}. On the contrary, tumor establishment in high-grade cell lines is repeatedly observed. Until now, the underlying mechanisms that lead to such different behaviors are unclear. We conjecture that the behavior of low-grade tumors resembles the attractive regime in our model while high-grade tumors behave as in the repulsive model regime. We suggest that the emergence of an Allee effect in low-grade tumors explains the existence of subcritical populations with low persistence probabilities. In contrast, the high-grade tumor cells always persists. Thus, we propose that the progression to malignancy may result from altered adaptions to the cellular microenvironment with respect to the regulation of cell migration and proliferation. 

The theoretical findings in our study might also provide suggestions for the design of new tumor therapies.  Standard tumor therapy, such as chemo-and radio-therapy, is directed towards controlling cell proliferation. However, a recent study found out that neoadjuvant chemotherapy selects for more migratory phenotypes at the expense of proliferative ones \cite{Almendro2014}. Our study shows that a possible therapeutic suggestion for malignant tumors is to combine conventional therapies with adjuvant treatments that restore sufficient contact inhibition of cell migration. In particular, if contact inhibition of migration is enforced, a sufficiently small tumor may die out due to the intrinsic cell population dynamics. On the contrary, if contact inhibition of cell migration is downregulated, any tumor inevitably grows and recurrence cannot be prevented. It is well known that malignant tumor cells lose sensitivity to contact inhibition of migration \cite{Abercrombie1979,Takai2008}. Our study shows that this is not only a bystander effect but a key determinant of tumor's fate. Further investigations are required for the experimental validation of our hypothesis.

\section*{Acknowledgments}
This work was supported by the Free State of Saxony and European Social Fund of the European Union (ESF, grant GlioMath-Dresden). MAH has been has been partially supported by MINECO Grant MTM2011-22656. HH acknowledges the support of the German Research Foundation (DFG) within the Cluster of Excellence 'Center for Advancing Electronics Dresden' and the German Ministry of Education and Research (BMBF) within the project SYSIMIT-01ZX1308D. We thank Ada Cavalcanti-Adam for helpful comments on the manuscript.


\section*{Figures}

\begin{thebibliography}{50}
	\bibitem{Klein2013}	Klein CA. Selection and adaptation during metastatic cancer progression. Nature 2013; 501:365-72.
	\bibitem{Meacham2013} Meacham CE, Morrison SJ. Tumour heterogeneity and cancer cell plasticity. Nature 2013; 501:328-37.
	\bibitem{Friedl2003} Friedl P, Wolf K. Tumour-cell invasion and migration: diversity and escape mechanisms.  Nat Rev Cancer 2003; 3:362-74.
	\bibitem{Friedl2011} Friedl P, Alexander S. Cancer invasion and the microenvironment: plasticity and peciprocity. Cell 2011; 147:992-1009.
	\bibitem{Cairns2011} Cairns RA, Harris IS, Mak TW. Regulation of cancer cell metabolism. Nat Rev Cancer 2011; 11:85-95.
	\bibitem{Gao2005} Gao CF, Xie Q, Su YL, Koeman J, Khoo SK, Gustafson M, Knudsen B, Hay R, Shinomiya N, {Vande Woude} GF. Proliferation and invasion: Plasticity in tumor cells. PNAS 2005; 120:10528-10533.
	\bibitem{DeDonatis2008} De Donatis A1, Comito G, Buricchi F, Vinci MC, Parenti A, Caselli A, Camici G, Manao G, Ramponi G, Cirri P. Proliferation versus migration in platelet-derived growth factor signaling: the key role of endocytosis. J Biol Chem 2008; 283:19948-56.	
	\bibitem{Zheng2009} Zheng P, Severijnen L, van~der Weiden M, Willemsen R, Kros JM. Cell proliferation and migration are mutually exclusive cellular phenomena in vivo: Implications for cancer therapeutic strategies. Cell Cycle 2009; 8:950-1.
	\bibitem{Farin2006} Farin A, Suzuki SO, Weiker M, Goldman JE, Bruce JN, Canoll P. Transplanted glioma cells migrate and proliferate on host brain vasculature: a dynamic analysis. Glia 2006; 53:799-808.
	\bibitem{Garay2013} Garay T, Juh\' asz \' E, Moln\' ar E, Eisenbauer M, Czir\' ok A, Dekan B, László V, Hoda MA, Döme B, Timár J, Klepetko W, Berger W, Hedegus B. Cell migration or cytokinesis and proliferation?--revisiting the ``go or grow'' hypothesis in cancer cells {\it in vitro}. Exp Cell Res 2013; 319:3094-103.
	\bibitem{Hoek2008} Hoek KS, Eichhoff OM, Schlegel NC, D\"{o}bbeling U, Kobert N, Schaerer L, Hemmi S, Dummer R. {\it In vivo} switching of human melanoma cells between proliferative and invasive states. Cancer Res 2008; 68:650-6.
	\bibitem{Jerby2012} Jerby L, Wolf L, Denkert C, Stein GY, Hilvo M, Oresic M, Geiger T, Ruppin E. Metabolic associations of reduced proliferation and oxidative stress in advanced breast cancer. Cancer Res 2012; 72:5712-20.
	\bibitem{Giese2003} Giese A, Bjerkvig R, Berens ME, Westphal M. Cost of migration: invasion of malignant gliomas and implications for treatment. J Clin Oncol 2003; 21:1624-36.
	\bibitem{Godlewski2010} Godlewski J, Bronisz A, Nowicki MO, Chiocca E A, Lawler S. microRNA-451: A conditional switch controlling glioma cell proliferation and migration. Cell Cycle 2010; 9:2742-8.
	\bibitem{Chauviere2010} Chauvi\`{e}re A, Preziosi L, Byrne H. A model of cell migration within the extracellular matrix based on a phenotypic switching mechanism. Math Med Biol 2010; 27:255-81.
	\bibitem{Hatzikirou2012} Hatzikirou H, Basanta D, Simon M, Schaller K, Deutsch A. 'Go or grow': the key to the emergence of invasion in tumour progression? Math Med Biol 2012; 29:49-65.
	\bibitem{Kim2011} Kim Y, Roh S, Lawler S, Friedman A. miR451 and AMPK mutual antagonism in glioma cell migration and proliferation: a mathematical model. PLoS One 2011; 6:e2829.
	\bibitem{Martinez2012} Mart\'{\i}nez-Gonz\'{a}lez A, Calvo GF, {P\'{e}rez Romasanta} LA , P\'{e}rez-Garc\'{\i}a VM. Hypoxic cell waves around necrotic cores in glioblastoma: a biomathematical model and its therapeutic implications. Bull Math Biol 2012; 74:2875-96.
	\bibitem{Pham2012} Pham K, Chauvi\`{e}re A, Hatzikirou H, Li X, Byrne HM, Cristini V, Lowengrub J. Density-dependent quiescence in glioma invasion: instability in a simple reaction-diffusion model for the migration/proliferation dichotomy. J Biol Dyn 2012; 6:54-71.
	\bibitem{Tektonidis2011} Tektonidis M, Hatzikirou H, Chauvi\`{e}re A, Simon M, Schaller K, Deutsch A. Identification of intrinsic {\it in vitro} cellular mechanisms for glioma invasion. J Theor Biol 2011; 287:131-47.
	\bibitem{Favaro2008} Favaro E, Nardo G, Persano L, Masiero M, Moserle L, Zamarchi R, Rossi E, Esposito  G, Plebani M, Sattler U, Mann T, Mueller-Klieser W, Ciminale V, Amadori A, Indraccolo S. Hypoxia inducible factor-1alpha inactivation unveils a link between tumor cell metabolism and hypoxia-induced cell death. Am J Pathol 2008; 173:1186-201.
\bibitem{Vultur2004} Vultur A, Cao J, Arulanandam R, Turkson J, Jove R, Greer P, Craig A, Elliot B, Raptis L. Cell-to-cell adhesion modulates Stat3 activity in normal and breast carcinoma cells. Oncogene 2004; 23:2600-16.
	\bibitem{Anderson2008} Anderson AR, Quaranta V. Integrative mathematical oncology. Nat Rev Cancer 2008; 8:227-34. 
	\bibitem{Byrne2010} Byrne HM. Dissecting cancer through mathematics: from the cell to the animal model. Nat Rev Cancer 2010; 10:221-30. 	
	\bibitem{Gatenby2003} Gatenby RA, Maini PK. Mathematical oncology: cancer summed up. Nature 2003; 421:321.
	\bibitem{Courchamp1999} Courchamp F, Clutton-Brock T, Grenfell B. Inverse density dependence and the Allee effect. Trends Ecol Evol 2012; 14:405-410.	
	\bibitem{Deutsch2005} Deutsch A, Dormann S. Cellular Automaton Modeling of Biological Pattern Formation. Birkh{\"a}user, 2005.
	\bibitem{Dormann2002} Dormann S, Deutsch A. Modeling of self-organized avascular tumor growth with a hybrid cellular automaton. In Silico Biol 2002; 2:393-40.
	\bibitem{Hatzikirou2008} Hatzikirou H, Deutsch A. Cellular automata as microscopic models of cell migration in heterogeneous environments. Curr Top Dev Biol 2008; 81:401-34.
	\bibitem{Taylor2005} Taylor CM, Hastings A. Allee effects in biological invasions. Ecol Lett 2005; 8:895-908.
	\bibitem{Mikhailov1994} Mikhailov AS. Foundations of synergetics I. Springer, 1994.
	\bibitem{Murray2003} Murray JD. Mathematical Biology II: spatial models and biomedical applications. Springer, 2003.
	\bibitem{Aronson1978} Aronson DG, Weinberger HF. Multidimensional nonlinear diffusion arising in population genetics. Adv Math 1978; 30:33-76.
	\bibitem{Hamel2012} Hamel F, Nadin G. Spreading properties and complex dynamics for monostable reaction-diffusion equations. Comm Part Diff Eq 2012; 37:511-537.
	\bibitem{Murray2002} Murray JD. Mathematical Biology: I. An Introduction. Springer, 2002
	\bibitem{Du2010} Du Y, Matano H. Convergence and sharp thresholds for propagation in nonlinear diffusion problems. J Eur Math Soc 2010; 12:279-312.
	\bibitem{Sanchez2010} Sanchez-Garduno F, Maini PK, Perez-Velazquez J. A non-linear degenerate equation for direct aggregation and traveling wave dynamics. Discrete Continuous Dyn Syst Ser B 2010; 13:455-487.
	\bibitem{Cherubini2012} Cherubini C, Gizzi A, Bertolaso M, Tambone V, Filippi S. A bistable field model of cancer dynamics. Commun Comput Phys 2012; 11:1-18.
	\bibitem{Dewhirst2009} Dewhirst S, Lutscher F. Dispersal in heterogeneous habitats: thresholds, spatial scales, and approximate rates of spread. Ecology 2009; 90:1338-45.
	\bibitem{Vergni2012} Vergni D, Iannaccone S, Berti S, Cencini M. Invasions in heterogeneous habitats in the presence of advection. J Theor Biol 2012;301:141-52.
	\bibitem{Huszthy2012} Huszthy PC, Daphu I, Niclou SP, Stieber D, Nigro JM, Sakariassen P\O , Miletic H, Thorsen F, Bjerkvig R. {\it In vivo} models of primary brain tumors: pitfalls and perspectives. Neuro Oncol 2012; 14(8):979-93.
	\bibitem{Tilkorn2011} Tilkorn D, Daigeler A, Stricker I, Schaffran A, Schmitz I, Steinstraesser L, Hauser J, Ring A, Steinau HU, Al-Benna S. Establishing efficient xenograft models with intrinsic vascularisation for growing primary human low-grade sarcomas. Anticancer Res 2011; 31:4061-6.	
	\bibitem{Almendro2014} Almendro V, Cheng YK, Randles A, Itzkovitz S, Marusyk A, Ametller E at al. Inference of tumor evolution during chemotherapy by computational modeling and in situ analysis of genetic and phenotypic cellular diversity. Cell Rep 2014; 6:514-527.	
	\bibitem{Abercrombie1979} Abercrombie M. Contact inhibition and malignancy. Nature 1979; 281:259-262.
	\bibitem{Takai2008} Takai Y, Miyoshi J, Ikeda W, Ogita H. Nectins and nectin-like molecules: roles in contact inhibition of cell movement and proliferation. Nat Rev Mol Cell Biol 2008; 9:603-15.	
\end{thebibliography}
\end{document}


\begin{flushleft}
	{\Large \textbf{Supplementary Information }} \\ \bigskip
\end{flushleft}

\begin{table}[h]
	\fbox{
	\normalsize
	\begin{tabular}{l p{0.65\textwidth}}
		\textbf{Symbol} & \textbf{Explanation}\\
		\hline
		\\
		$\mathscr{L}\subset\mathbb{Z}^d $ & d-dimensional regular square lattice  \\[0.8ex]
		$\br\in\mathscr{L} $ & node on the lattice  \\[0.8ex]
		$\sigma \in\{m,r\}$ & cell type in the model (moving $m$ and resting $r$)  \\[0.8ex]
		$K\geq 2d$ & capacity of each lattice node ($2d$ velocity channels and $K-2d$ resting channels) \\[0.8ex]
		$n (\br,k) \in \{0,\dots,K\}$ & cell number at a node $\br$ and time $k$  \\[0.8ex]
		$\cd (\br,k) = n(\br,k)/K $ & cell density at a node $\br$ and time $k$ ($0\leq\cd (\br,k)\leq 1$)  \\[0.8ex]
		$\mcd (\br,k) = \langle n(\br,k) \rangle /K  $ & mean cell density  at a node $\br$ and time $k$ ($0\leq\mcd (\br,k)\leq 1$) \\[0.8ex]
		$k\in \mathbb{N}$ & automaton time step  \\[0.8ex]
		$r_d$ & probability that a cell dies \\[0.8ex]
		$r_b$ & probability that a resting cell undergoes mitosis \\[0.8ex]
		$r_s$ & probability that a moving cell changes to resting  \\[0.8ex]
		$1-r_s$ & probability that a resting cell changes to moving  \\[0.8ex]
		$\kappa \in \mathbb{R}$ & intensity of the phenotypic switch \\[0.8ex]
		$\theta \in (0,1)$ & critical cell density value at which the probabilities to switch from resting to moving and vice versa are equal \\[0.8ex]
		$\Prob(\cdot)$ & probability \\[0.8ex]
		$T_b$ & average cell cycle time  \\[0.8ex]
		$R_b$ & real birth rate ($T_b=1/R_b$) \\[0.8ex]
		$T_d$ & average life time of a cell \\[0.8ex]
		$R_d$ & real death rate ($T_d=1/R_d$) 
	\end{tabular}}
	\caption{List of symbols}
	\label{tab:symbols}
\end{table}

\section{LGCA model description}
\label{sec:model}
Our LGCA is defined on a 2-dimensional square lattice $\mathscr{L}\subset\mathbb{Z}^2$. To every lattice node $\br\in\mathscr{L}$, velocity channels $(\br,\fc_i)$, $i=1,\dots,b$, are associated. The parameter $b$ defines the number of nearest neighbors on the lattice. Here, we choose $b = 4$ and $c_i\in \{(1,0),(0,1),(-1,0),(0,-1)\}$. In addition, a fixed number of rest channels, $(\br,\fc_i)$, $i=b+1,\dots,K$ with $c_i=\{(0,0)\}$ is introduced. The LGCA imposes an exclusion principle on channel occupation, i.e.\ at any time at most one cell is allowed in each channel at every lattice node. Thus $K$ defines the maximal capacity of cells per node. We represent the healthy cells by empty channels and model explicitly two tumor cell phenotypes, denoted by $\sigma \in \{m,r\}$: moving ($m$) and resting ($r$) tumor cells. The occupation numbers at time $k$, $\eta_i(\br,k)$, $i=1,\dots,K$, are random Boolean variables that indicate the presence ($\eta _i=1$) or absence ($\eta _i=0$) of a tumor cell in the channel $(\br,\fc_i)$. The local configuration of cells at a node $\br$ at time $k$ is described by a vector $\boldsymbol{\eta}(\br,k)=(\eta_1(\br,k),\dots,\eta_K(\br,k))\in\{0,1\}^K$. The total number of tumor cells at a node $\br$ and time $k$ is defined by 
$$n (\br,k)= n_m(\br,k) + n_r(\br,k) = \sum\limits _{i=1}^{b} \eta _i (\br,k) + \sum\limits _{i=b+1}^{K} \eta _i (\br,k),$$
where $n_m(\br,k)$ and $n_r(\br,k)$ are the number of moving and resting cells at a node $\br$ at time $k$, respectively.

The dynamics of our LGCA arise from the repeated application of three operators: {\bf  propagation} ($\mathcal{P}$), {\bf reorientation} ($\mathcal{O}$) and {\bf cell reactions} ($\mathcal{R}$). The propagation and reorientation operators define cell movement, whereas the cell reactions operator changes the local number of cells on a node. The composition $\mathcal{R}\circ \mathcal{O} \circ \mathcal{P}$  of the three operators is applied independently at every node $\br$ of the lattice and at each time $k$ to evaluate the configuration at time $k+1$. 

The cell reactions operator comprises cell death and cell proliferation where the latter is affected by the migration-proliferation dichotomy. In detail, we consider three stochastic processes: phenotypic switching ($\RS$), cell death($\RD$) and proliferation($\RB$). For the sake of simplicity, the interaction between cells is restricted to the node itself. Moreover, we consider interaction rules which are independent of the cells' distribution to channels and solely depend on the total number of cells $n(\br,k)$ within the node. During one automaton time step, the number of cells on a node $\br$ changes by subsequent application of a switch ($\RS$), death ($\RD$) and proliferation ($\RB$) step,
\begin{equation} \label{eq:kinetics}
	n=(n_m,n_r) \xrightarrow{\RS} (n_m',n_r') \xrightarrow{\RD} (n_m'',n_r'') \xrightarrow{\RB} (n_m'',n_r''')=n^{\mathcal{R}}.
\end{equation}
The precise update rules are described below. Since we assume that individual decide to cells switch, die or divide independently from each other, the corresponding transition probabilities for the cell number per node follow a binomial distribution.
\begin{itemize}
	\item {\it Phenotypic switch ($\RS$)}: Cells can either rest or move. The phenotypic switching depends on the local cell density $\varrho=n/K$. Phenotypes are changed with probabilities $r_s(\varrho)$  and $1-r_s(\varrho)$ that denote the probabilities of a moving cell to become resting and vice versa. Moving and resting populations can exchange cells if there is enough free space on the lattice. Two steps are independently performed:
	\begin{itemize}
		\item[(1)] Define $M_1=\mathrm{min}(n_m,(K-b)-n_r)$ which is the potential number of cells that can switch from the moving to the resting phenotype, taking into account the effect of local volume exclusion. The transition probability that there are $j_1$ successful events `moving$\rightarrow$resting' is modeled by
		\begin{equation} \label{eq:binomCellSwitch1}
			P_{\RS_1}((n_m,n_r)\rightarrow (n_m-j_1,n_r+j_1))=
			\begin{cases}
				\binom{M_1}{j_1}r_s(\varrho)^{j_1}(1-r_s(\varrho))^{M_1-j_1} & \mbox{ if } 0\leq j_1\leq M_1 , \\
				0 & \mbox{ else}.
			\end{cases}
		\end{equation}	
		\item[(2)] Define $M_2=\mathrm{min}(n_r,b-n_m)$ which is the potential number of cells that can switch from the moving to the resting phenotype, taking into account the effect of local volume exclusion. The transition probability that there are $j_2$ successful events `resting$\rightarrow$moving' is given by	 
		\begin{equation} \label{eq:binomCellSwitch2}
			P_{\RS_2}((n_m,n_r)\rightarrow (n_m+j_2,n_r-j_2))=
			\begin{cases}
				\binom{M_2}{j_2}(1-r_s(\varrho))^{j_2}r_s(\varrho)^{M_2-j_2} & \mbox{ if } 0\leq j_2\leq M_2 , \\
				0 & \mbox{ else}.
			\end{cases}
		\end{equation}	
	\end{itemize}
	Finally, the difference between successful switches `moving$\rightarrow$resting' and `resting$\rightarrow$moving' is used to update the number of moving and resting cells so that $(n_m,n_r)\rightarrow (n_m-(j_1-j_2),n_r+(j_1-j_2))=(n_m',n_r')$. Note that the switching process does not change the total number of cells but the ratio of moving and resting cells.	
	
	\item {\it Cell death ($\RD$)}: Each tumor cell is assumed to die with probability $r_d$, independently from the other cells. Thus, the probability that $j_1$ moving and $j_2$ resting cells die, starting with $n'$ cells is 	
\end{itemize}	
\begin{equation} \label{eq:binomCellDeath}
	P_\RD((n_m',n_r')\rightarrow (n_m'-j_1,n_r'-j_2))=
	\begin{cases}
		\binom{n'}{j}r_d^{j}(1-r_d)^{n'-j} & \mbox{ if } j_1\leq n_m' \mbox{ and } j_2\leq n_r',\ j=j_1+j_2 , \\
		0 & \mbox{ else}.
	\end{cases}
\end{equation}	
\begin{itemize}
	\item {\it Cell proliferation ($\RB$)}: Birth of cells is dictated by the migration-proliferation dichotomy, which means that cells are allowed to proliferate only when they rest, i.e.\ when they are positioned on a rest channel. In addition, there must be empty rest channels to place the offspring into. Therefore, $M=\mathrm{min}(n_r'',(K-b)-n_r'')$ defines the potential number of resting cells that are allowed to proliferate starting from $n''$ cells after the switch and death processes. The transition probability that there are $j$ new cells after the proliferation step is then given by
	\begin{equation} \label{eq:binomCellBirth}
		P_\RB((n_m'',n_r'')\rightarrow (n_m'',n_r''+j))=
		\begin{cases}
			\binom{M}{j}r_b^{j}(1-r_b)^{M-j} & \mbox{ if } j'\leq M , \\
			0 & \mbox{ else}.
		\end{cases}
	\end{equation}	
\end{itemize}
Altogether changes occurring in the cell number of a given node between two consecutive times $k$ and $k+1$ can be encoded in a transition matrix $P$ with
\begin{equation}
	P = P_\RS P_\RD P_\RB, \label{eq:transMat}
\end{equation}
where the entries of the transition matrix of the switch process $P_\RS$ is determined by \eqref{eq:binomCellSwitch1} and \eqref{eq:binomCellSwitch2}, and the transition matrices $P_\RD$ and $P_\RB$ of the death and proliferation processes, respectively, are defined by \eqref{eq:binomCellDeath} and \eqref{eq:binomCellBirth}.

Subsequent to the cell reactions step, a reorientation step takes place where cells are randomly redistributed to the channels, providing a new local cell configuration at each node. Because of the migration-proliferation dichotomy, only the cells of migratory phenotype move. Therefore, moving cells are distributed to the velocity channels, while resting cells are distributed to the resting channels, resulting in a configuration $\eta^{\mathcal{R}\circ \mathcal{O}}(\br)$.

The final update step is the application of a propagation operator. Cells in the velocity channels, i.e.\ moving cells, are transported one lattice unit in directions determined by their velocities. The movement of cells in the propagation step is purely deterministic. The spatio-temporal automaton dynamics is therefore completely described by the mircodynamical equation
\begin{equation} \label{eq:micro}
	\eta _i(\br+\fc_i,k+1)-\eta _i(\br,k) = \eta _i^{\mathcal{R}\circ \mathcal{O}}(\br,k)-\eta _i(\br,k) \in  \{-1,0,1\} , \ \br\in\mathcal{L},\ i=1,\dots,K,\ k=0,1,\dots,
\end{equation}
where the change in the occupation numbers is $-1$ if a channel looses a cell, $0$ if nothing is changed and $1$ if a channel gains a cell, respectively.

\section{Scaling of the LGCA}
\label{sec:scaling}
We have chosen a nondimensional scaling with unit lattice spacing and unit time scale $k\in\mathbb{N}$ in our LGCA simulation. Thus, cells residing at velocity channels move one lattice unit per unit time step. Our dimensionless simulations can easily be rescaled such that the temporal and spatial scales fit to specific applications. In our case, cell death and proliferation probabilities are related to the real dimensional average life time $T_d$ of a cell and the average cell cycle time $T_b$, respectively, by
\begin{equation}
	r_d(\tau)=\frac{\tau}{T_d},\ r_b(\tau)=\frac{\tau}{T_b}, \label{eq:deathScaling}
\end{equation}
where $\tau$ is a sufficient small real dimensional time step length. This scaling is chosen because of the following arguments, detailed for the case of cell death. Let $X_1,X_2,\dots$ be independent Bernoulli trials with possible outcomes ``cell dies'' ($X_i=1$) and ``cell survives'' ($X_i=0$). The corresponding probability distribution is given by $\Prob(X_i=1)=r_d$. Let $N$ be the number of trials up to the first time a cell dies, i.e. $N:=\min\{n|X_1=\dots X_{n-1}=0,X_n=1\}$. Then $N$ is geometrically distributed such that $\Prob(N=n)=(1-r_d)^{n-1}r_d$ with expected value $\mathbb{E}[N]=1/r_d$. If $\tau$ is the real dimensional time step length, the life time of a cell is given by $\tau N$. Hence, the average cell life time in the model is $T=\mathbb{E}[\tau N]=\tau/r_d$ which must be related to the real dimensional average life time $T_d$.

Similar arguments hold for the real macroscopic time taken for a cell to proliferate (cell cycle time) and for a cell to change its phenotype. Please note that if the LGCA parameters $r_s$, $r_d$ and $r_b$ depend on time $\tau$, the transition matrix of the the cell reactions becomes time-dependent: 
\begin{equation}
	P=P(\tau)=P_\RS(\tau) P_\RD(\tau) P_\RB(\tau). \label{eq:transMat2}
\end{equation}

Besides scaling the LGCA time intervals, also the lattice spacing can be adjusted. If the lattice spacing is $\epsilon$, the mean square displacement of a cell per time step $\tau$ in the LGCA is proportional to $\epsilon^2/\tau$. Thus, by scaling $\mathbf{x}=\epsilon \br\in\mathbb{R}$, the relationship between the LGCA motility and measured movement of cells is characterized by the diffusion constant $D=\epsilon^2/\tau$.

\section{Equilibrium state under fast switching assumption}
\label{sec:switchEqui}
When deriving the mean-field equation, we assume that the switch dynamics is much faster than the cell number changes due to proliferation and cell death, that is $r_s, 1-r_s \gg r_b,r_d$. Further, we consider the low density regime where the carrying-capacity effects can be neglected when considering the switching. Then, by time scale separation, the fractions $p_\sigma:= n_\sigma/n,\,\sigma \in \{r,m\}$ of resting and moving cells, respectively, are in detailed balance with respect to the switch dynamics. This amounts to 
\begin{align*}
    p_r (1-r_s) & = p_m r_s\\
                & = (1-p_r) r_s.
\end{align*}
Hence $  p_r = r_s $ or, equivalently, 
\begin{equation} \label{eq:EquiSwi}
	n_r= r_s n.
\end{equation}

\section{Mean-field approximation of the LGCA dynamics}
\label{sec:meanField}
The LGCA model is composed of a cell movement and a cell reaction process. In the following, we separate the two processes and derive mean-field approximations for each one. We show that the cell movement in the automaton model can be approximated by a degenerate diffusion term, while cell reaction in the automaton leads to a nonlinear reaction term in the resulting partial differential equation description.

\bigskip
\subsection{Scaling limit of the cell reaction process}
\label{sec:bdProcess}
In the following, we are interested in the time evolution of the cell density. For that reason, we neglect any spatial dependence between nodes. We assume that the system is in equilibrium with respect to the switching (fast switching assumption, see section \ref{sec:switchEqui}). We consider the the Markov process $(N_k)_{k\geq 0}$ which describes the number of cells at a node $r\in\mathcal{L}$ in the discrete phase space $\{0,\dots,K\}$. The transition matrix of this process, derived from \eqref{eq:transMat} with utilization of \eqref{eq:EquiSwi}, is given by $ W=W_\RD W_\RB$, where
\begin{eqnarray}
	W_\RD(i,j)  &=& \begin{cases}
				\binom{i}{j}(1-r_d)^{j}r_d^{i-j}   & \mbox{ if } j\leq i ,\\
				0 & \mbox{ else}
			\end{cases} \label{eq:TransMatrixD}
\end{eqnarray}
and 
\begin{eqnarray}
	W_\RB(i,j)  &=& \begin{cases}
				\binom{M}{r_sj-r_si}r_b^{r_sj-r_si}(1-r_b)^{M-r_sj+r_si}  & \mbox{ if } r_sj-r_si\leq M=\mathrm{min}(r_si,K-i) ,\\
				0 & \mbox{ else},
			\end{cases} \label{eq:TransMatrixB}
\end{eqnarray}
Here, the minimum function $\mathrm{min}(x,K-x)\approx x(K-x) \frac{1}{K}$, $x\in(0,K)$. This approximation underestimates the minimum function but provides an intuitive explanation as it describes the fraction of cells that are able to proliferate times the fraction of space that can be filled. Please note that, for simplicity, we use a notation where the dependence of the switch function $r_s$ from the node density $\cd$ is dropped.

The probability to be in state $j$ after one time step is 
$$ \Prob_{k+1}(j):=\Prob(N_{k+1}=j)=\sum\limits_{i=0}^{K}W(i,j) \Prob(N_k=i)=\sum\limits_{i=0}^{K}W(i,j) \Prob_k(i)$$
and in matrix notation
$$ \Prob_{k+1}=W\Prob_k.$$
The transition of a microscopic process to a macroscopic description requires a temporal scaling relation between the macroscopic and microscopic variables (see section \ref{sec:scaling} for details). We assume a small parameter $\tau>0$ that scales the time variable $t=k\tau$ such that
$$ \Prob_{t+\tau} = W(\tau)\Prob_t ,$$
where now the transition matrix $W$ depends on the real dimensional time step size $\tau$. Rearranging terms gives
\begin{eqnarray*}
	\Prob_{t+\tau}-\Prob_t &=&  (W(\tau)-I)\Prob_t ,\\
	\frac{\Prob_{t+\tau}-\Prob_t}{\tau} &=& \frac{W(\tau)-I}{\tau}\Prob_t .
\end{eqnarray*}
Then, for $\tau \downarrow 0$
\begin{equation}
	\frac{\partial}{\partial t} \Prob_t	= Q \Prob_t \label{eq:master}
\end{equation}
with intensity matrix $Q:= \lim\limits_{\tau \downarrow 0}\frac{W(\tau)-I}{\tau} = \left. \frac{\partial}{\partial \tau} W(\tau) \right|_{\tau=0}$. Since $W(\tau)=W_\RD(\tau) W_\RB(\tau)$ it follows that
\begin{eqnarray}
	Q = \left. \frac{\partial}{\partial \tau} W(\tau) \right|_{\tau=0} &=& \left.\left[ \left(\frac{\partial}{\partial \tau} W_\RD(\tau)\right)W_\RB(\tau) + W_\RD(\tau) \left(\frac{\partial}{\partial \tau} W_\RB(\tau)\right) \right] \right|_{\tau=0} \nonumber \\
	&=& \left. \frac{\partial}{\partial \tau} W_\RD(\tau) \right|_{\tau=0} + \left. \frac{\partial}{\partial \tau} W_\RB(\tau) \right|_{\tau=0}  \nonumber \\
	&=& Q_\RD + Q_\RB,  \label{eq:TransMatrixLimit}
\end{eqnarray}
where $Q_\RD$ and $Q_\RB$ are the intensity matrices for the death and birth processes, respectively. A direct calculation gives
\begin{eqnarray}
	Q_\RD = \lim\limits_{\tau \downarrow 0} \frac{1}{\tau} W_\RD(\tau;i,j) &=& \lim\limits_{\tau \downarrow 0} \frac{1}{\tau}
	\begin{cases}
		\binom{i}{j}(1-r_d(\tau))^{j}r_d(\tau)^{i-j} & \mbox{ if } 0\leq j\leq i ,\\
		0 & \mbox{ else},
	\end{cases} \nonumber \\
	& \stackrel{\eqref{eq:deathScaling}}{=} & \lim\limits_{\tau \downarrow 0} \frac{1}{\tau}
	\begin{cases}
		\binom{i}{j}\left(1-\frac{\tau}{T_d}\right)^{j}\left(\frac{\tau}{T_d}\right)^{i-j} & \mbox{ if } 0\leq j\leq i ,\\
		0 & \mbox{ else},
	\end{cases} \nonumber \\
	&=& \lim\limits_{\tau \downarrow 0} 
	\begin{cases}
		\frac{1}{T_d}\binom{i}{l}\left(1-\frac{\tau}{T_d}\right)^{j}\left(\frac{\tau}{T_d}\right)^{i-j} & \mbox{ if } 0\leq j\leq i ,\\
		0 & \mbox{ else},
	\end{cases} \nonumber \\
	&=& \begin{cases}
		\frac{i}{T_d} & \mbox{ if } j=i-1 ,\\
		0 & \mbox{ else}
	\end{cases} \label{eq:TransMatrixLimitDeath}
\end{eqnarray}
and similarly
\begin{equation}
	Q_\RB = \lim\limits_{\tau \downarrow 0} \frac{1}{\tau} W_\RB(\tau;i,j) = \begin{cases}
		\frac{1}{T_b}r_si\left(1-\frac{i}{K}\right) & \mbox{ if } r_sj=r_si-1 ,\\
		0 & \mbox{ else}.
	\end{cases} \label{eq:TransMatrixLimitBirth}
\end{equation}
From the master equation \eqref{eq:master}, the temporal evolution of the mean cell density $\mcd_t=\langle N_t\rangle/K$ is obtained by
\begin{eqnarray}
	 \frac{\partial}{\partial t} \mcd_t \ =\  \frac{1}{K}\sum\limits_{j=0}^{K} j \frac{\partial}{\partial t} \Prob_t(j) &=&  \frac{1}{K} \sum\limits_{j=0}^{K} j \sum\limits_{i=0}^{K}Q(i,j) \Prob_t(i) . \\
	 &=& \frac{1}{K} \sum\limits_{i=0}^{K}  \Prob_t(i)  \sum\limits_{j=0}^{K} j Q(i,j) \nonumber \\
	 & \stackrel{\eqref{eq:TransMatrixLimit}}{=} & \frac{1}{K} \sum\limits_{i=0}^{K}  \Prob_t(i)  \sum\limits_{j=0}^{K} j Q_\RD(i,j) +  \frac{1}{K} \sum\limits_{i=0}^{K}  \Prob_t(i)  \sum\limits_{j=0}^{K} j Q_\RB(i,j) \nonumber \\
	  & \stackrel{\eqref{eq:TransMatrixLimitDeath},\eqref{eq:TransMatrixLimitBirth}}{=} &  \frac{1}{K} \frac{1}{T_d} \sum\limits_{i=0}^{K}  (-i)  \Prob_t(i)+ \frac{1}{K} \frac{1}{T_b} \sum\limits_{i=0}^{K}  r_s i\left(1- \frac{i}{K}\right) \Prob_t(i) \nonumber \\
	  &=& -\frac{1}{T_d}\mcd + r_s \frac{1}{T_b}\mcd  - r_s\frac{1}{K} \frac{1}{T_b} \sum\limits_{i=0}^{K} i^2 \Prob_t(i)
\end{eqnarray}
Assuming that $\mathrm{E}[N_t^2]=\mathrm{E}[N_t]^2$, which holds if the variance of $N_t$ is small, for instance if $K$ is large, we finally arrive at a macroscopic description of the LGCA birth-death process
\begin{equation}
	\frac{\partial}{\partial t} \mcd_t =  -R_d\mcd + r_sR_b\mcd(1-\mcd) , \label{eq:BDProcess}
\end{equation}
where $R_d=1/T_d$ and $R_b=1/T_b$ are the real death and proliferation rates, respectively, see section \ref{sec:scaling}.

\subsection{Scaling limit of the cell migration process}
\label{sec:jumpProcess}
In the following, we consider the automaton cell migration process without cell reaction, under the fast switching assumption, see section \ref{sec:switchEqui}. In our LGCA, the probability of a cell to jump from one node $\br$ to a neighboring node $\br+\fc_i$ is equal to the probability of a cell residing in the respective velocity channel. In the isotropic case, cells within velocity channels are redistributed with probability $1/b$. To connect the discrete mechanism with a continuum model we consider the expected cell density of a node $r$ at time $k$, defined by $\mcd(r,k)=\langle\cd(r,k)\rangle$, in the scaling limit when time step length and lattice spacing goes to 0. 

The change in average occupancy of a moving cell at site $r$ during the next time step is given by
\begin{dmath} \label{eq:jump}
	\mcd(\br,k+1) = \sum_{i=1}^b \frac{1-r_s(\mcd(\br+\fc_i,k))}{b}\mcd(\br+\fc_i,k)  + r_s(\mcd(\br,k))\mcd(\br,k),
\end{dmath}
where the terms can be interpreted as the probability that a cell at site $\br+\fc_i$, $i=1,\dots,b$, is moving to site $\br$ and the probability that a moving cell at site $\br$ does not attempt to leave the site. In the following, for simplicity, we will analyze a one-dimensional motility mechanism where motility events only take place in the horizontal direction. This approach can easily be extended to higher dimensions, because of the model isotropy. 

In one dimension, equation \eqref{eq:jump} reduces to 
\begin{dmath} \label{eq:jump1}
	\mcd(\br,k+1) =  \frac{1-r_s(\mcd(\br-1,k))}{2}\mcd(\br-1,k) + \frac{1-r_s(\mcd(\br+1,k))}{2}\mcd(\br+1,k)  + r_s(\mcd(\br,k))\mcd(\br,k),
\end{dmath}
where the terms correspond to the probability that a cell at site $\br-1$ is moving and moves to the right, the probability that a a cell at site $\br+1$ is moving and moves to the left and the probability that a moving cell at site $\br$ does not attempt to leave the site.
Approximation of $r_s(\mcd)$ around $\mcd \ll 1 $ such that $r_s(\mcd)=r_s(0)+ r_s'(0)\mcd+r_s''(0)\mcd^2 +\mathcal{O}(\mcd^3)$, equation \eqref{eq:jump1} becomes
\begin{dmath} \label{eq:jump2}
	\mcd( \br,k+1)	= \frac{1}{2} \left(\mcd(\br+1,k) + \mcd(\br-1,k) \right)  - \frac{1}{2} r_s(0)(\mcd(\br+1,k) + \mcd(\br-1,k)) - \frac{1}{2}  r_s'(0)(\mcd^2(\br+1,k) + \mcd^2(\br-1,k))- \frac{1}{2} r_s''(0)(\mcd^3(\br+1,k) + \mcd^3(\br-1,k))  + r_s(0)\mcd(\br,k) +  r_s'(0)\mcd^2(\br,k) + r_s''(0)\mcd^3(\br,k) \nonumber
\end{dmath}
To convert \eqref{eq:jump2} into a continuous macroscopic partial differential equation (PDE), we identify the average density $\mcd(\br,k)$ by its continuous counterpart $\mcd(\fx,t)$, where $\fx= \br\epsilon\in\mathbb{R}$ and $t=k \tau\in\mathbb{R}_{+}$, with $\epsilon,\tau\in\mathbb{R}_+$, which gives 
\begin{dmath} \label{eq:jump3}
	\mcd(\fx ,t+\tau )	= \frac{1}{2} \left(\mcd(\fx+\epsilon,t) + \mcd(\fx-\epsilon,t) \right) - \frac{1}{2} r_s(0)(\mcd(\fx+\epsilon,t) + \mcd(\fx-\epsilon,t)) - \frac{1}{2}  r_s'(0)(\mcd^2(\fx+\epsilon,t) + \mcd^2(\fx-\epsilon,t)) - \frac{1}{2} r_s''(0)(\mcd^3(\fx+\epsilon,t) + \mcd^3(\fx-\epsilon,t)) + r_s(0)\mcd(\fx,t) +  r_s'(0)\mcd^2(\fx,t) + r_s''(0) \mcd^3(\fx,t)  ,
\end{dmath}
Expanding all terms in \eqref{eq:jump3} in a truncated Taylor series in powers of $\epsilon$ and $\tau$ up to second order gives
\begin{dmath} \label{eq:jump4}
 	\mcd + \tau \partial_t \mcd + \frac{\tau^2}{2}\partial_{tt}\mcd = \mcd + \frac{\epsilon^2}{2}\partial_{xx} \mcd  -r_s(0)\mcd - \frac{r_s(0)\epsilon^2}{2}\partial_{xx} \mcd -  r_s'(0)\mcd^2 -  r_s'(0)\epsilon^2(\partial_{x} \mcd)^2 -  r_s'(0)\mcd\epsilon^2\partial_{xx} \mcd  - \frac{ r_s'(0)\epsilon^4}{2}(\partial_{xx} \mcd)^2  -r_s''(0)\mcd^3-3r_s''(0)\mcd\epsilon^2(\partial_{x} \mcd)^2-\frac{3r_s''(0)\epsilon^2}{2}\mcd^2\partial_{xx} \mcd-\frac{3r_s''(0)\epsilon^4}{2}(\partial_{x} \mcd)^2\partial_{xx} \mcd  - \frac{3r_s''(0)\epsilon^4}{4}\mcd (\partial_{xx} \mcd)^2  - \frac{r_s''(0)\epsilon^4}{2}(\partial_{xx} \mcd)^3 + r_s(0)\mcd +  r_s'(0)\mcd^2 + r_s''(0)\mcd^3  ,
\end{dmath}
To proceed, we consider the diffusive limit, i.e. $\epsilon\rightarrow 0$, $\tau\rightarrow 0$ and $\lim\limits_{\epsilon,\tau \rightarrow 0}{\frac{\epsilon^2}{\tau}}=const:=D$. This gives
\begin{dmath} \label{eq:jump5}
	\partial_t \mcd  = \frac{D}{2}  - \frac{D}{2}r_s(0) \partial_{xx} \mcd  - r_s'(0)D  (\partial_{x} \mcd)^2 - r_s'(0)D\mcd \partial_{xx} \mcd -3r_s''(0)D\mcd (\partial_{x} \mcd)^2 - \frac{3D}{2}r_s''(0)\mcd^2 \partial_{xx} \mcd  \\
	= D \partial_{xx}\mcd \underbrace{  \left( \frac{1-r_s(0)}{2} -  r_s'(0)\mcd -\frac{3}{2}r_s''(0)\mcd^2 \right) }_{=:\mathcal{D}(\mcd)} + D (\partial_{x} \mcd)^2 \underbrace{   \left( - r_s'(0)-3r_s''(0)\mcd \right) }_{=\mathcal{D}'(\mcd)}
\end{dmath}
Finally, the jump process of our LGCA can be approximated by a degenerate diffusion equation
\begin{equation} \label{eq:DriftDiffusion}
	\partial_t \mcd = D \partial_x (\mathcal{D}(\mcd) \partial_x \mcd) \ .
\end{equation}

\section{Kernel density estimation of the averaged cell density}
\label{sec:kde}
The kernel density estimation, also known as Parzen-Rosenblatt density estimation, is a well known nonparametric way of estimating the probability density function underlying a finite set of observations \cite{Rosenblatt,Parzen}. We use a built-in Mathematica function to estimate the kernel density of the averaged cell density. For details see the Wolfram Demonstrations Project \url{http://demonstrations.wolfram.com/KernelDensityEstimation/} and \url{http://reference.wolfram.com/mathematica/ref/SmoothKernelDistribution.html}.

\section{Stability analysis of cell reaction mean-field equation}
\label{sec:stabilityAnalysis}
In the following we analyze the stability behavior of the macroscopic growth equation
\begin{equation}\label{eq:stability}
	\partial_t \mcd =  F(\mcd) ,
\end{equation}
with
\begin{equation}\label{eq:stability2}
	F(\mcd)	= r_s(\mcd)r_b\mcd (1-\mcd) - r_d \mcd ,
\end{equation}
where 
$0<r_d\ll r_b\ll 1$ and $ r_s(\varrho) = \frac{1}{2}(1+\tanh (\kappa (\varrho-\theta)))$, $\mcd\in [0,1]$, 
$\theta\in(0,1)$ and $\kappa\in\mathbb{R}$.

Obviously, $\mcd_1^*=0$ is a fixed point of \eqref{eq:stability} and
$$ F'(0)= \frac{r_b}{2}(1-\tanh(\kappa\theta)) -r_d ,$$
where
\begin{equation}\label{eq:stabilityCases1}
	F'(0) 
	\begin{cases}
		>0 & \mbox{ if } \kappa<0 , \\
		>0 & \mbox{ if } \kappa>0 \mbox{ and } 1-2\frac{r_d}{r_b}<\tanh(\kappa\theta), \\
		<0 & \mbox{ if } \kappa>0 \mbox{ and } 1-2\frac{r_d}{r_b}>\tanh(\kappa\theta).
		\end{cases}
\end{equation}
Thus, the fixed point $\mcd_1^*=0$ is always unstable in the repulsive case ($\kappa>0$). In the attractive case ($\kappa>0$), $\mcd_1^*=0$ is stable if and only if $1-2\frac{r_d}{r_b}>\tanh(\kappa\theta)$ but unstable if $1-2\frac{r_d}{r_b}<\tanh(\kappa\theta)$.

\begin{figure}[HT]
	\centering
	\includegraphics[width=0.9\textwidth]{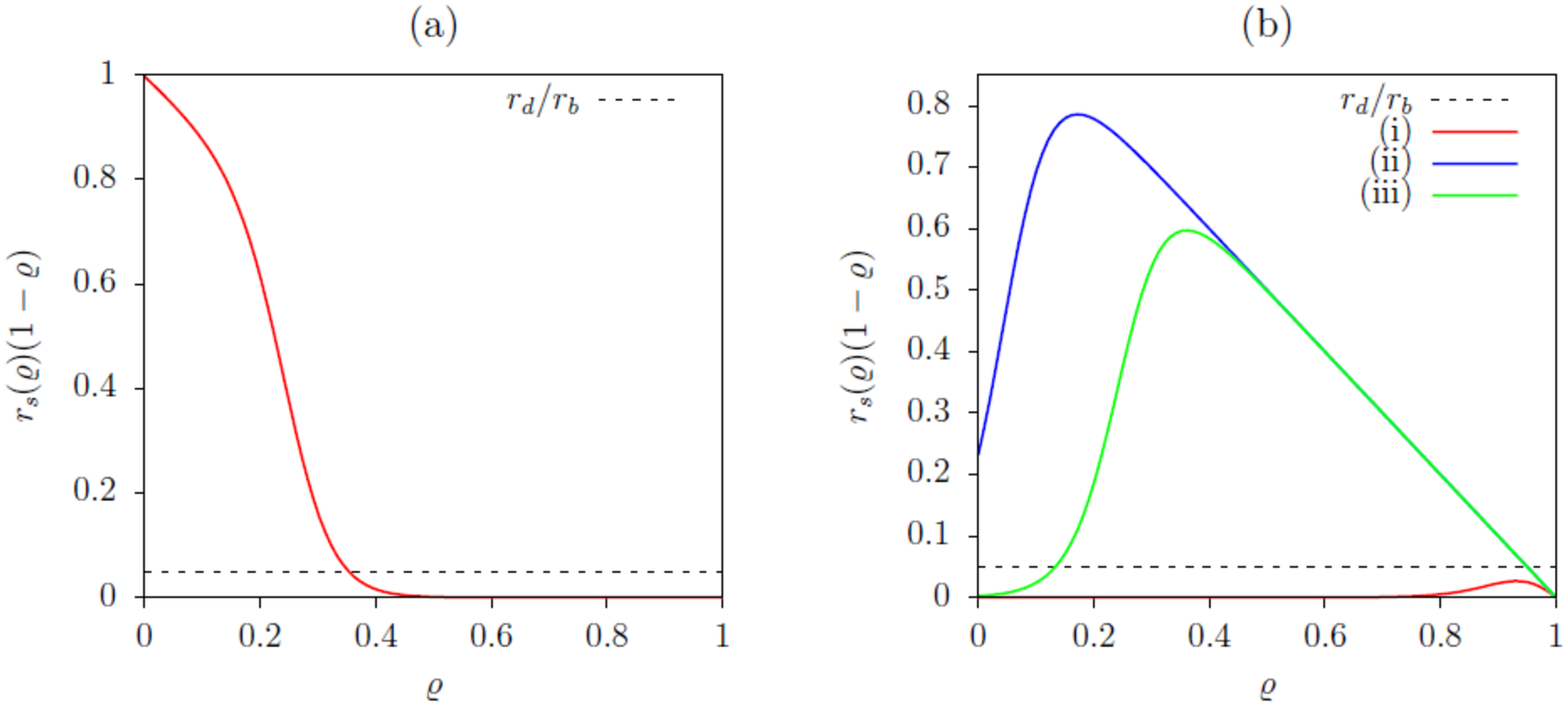}
	\caption{Stability analysis of the per capita growth rate function $f$ defined in \eqref{eq:stability3}. The dotted line represents the constant value $r_d/r_b$. The colored lines represent the function $r_s(\mcd)(1-\mcd)$ for different values of $\kappa$ and $\theta$. Intersection of both graphs indicate zeros of the function $f$ which correspond to fixed points of the LGCA cell reaction mean-field equation \eqref{eq:stability}. The sign of the slope of function $r_s(\mcd)(1-\mcd)$ at the intersection point reveals the stability behavior of the equivalent fixed point. (a) In the repulsive case ($\kappa<0$), there is exactly one intersection point. Parameters are $\kappa=-12$ and $\theta=0.25$. (b) In the attractive case ($\kappa>0$), three cases can be distinguished: (i) no  intersection point (red line, $\kappa=12$, $\theta=0.95$), (ii) one intersection point (blue line, $\kappa=12$, $\theta=0.05$) or (iii) two (green line, $\kappa=12$, $\theta=0.25$) intersection points between $r_d/r_b$ and $r_s(\mcd)(1-\mcd)$. }
	\label{fig-S1}
\end{figure}

Now, for $\mcd>0$ we define
\begin{equation}\label{eq:stability3}
	f(\mcd):= r_b r_s(\mcd)(1-\mcd) - r_d,
\end{equation}
which is the per capita growth rate of $F$. Fixed points $\mcd^*>0$ of equation \eqref{eq:stability} are equivalent to zeros of the function $f(\mcd)=0$, that is if they satisfy
\begin{equation}\label{eq:stability4}
	r_s(\mcd)(1-\mcd)= \frac{r_d}{r_b}.
\end{equation}
In addition, for fixed points $\mcd^*$ of equation \eqref{eq:stability} it holds that 
\begin{eqnarray}
	F'(\mcd^*)
	&=&  r_br_s(\mcd^*)(1-\mcd^*)-r_b\mcd^* r_s(\mcd^*)+r_b\mcd^*(1-\mcd^*)r_s'(\mcd^*) -r_d  \nonumber \\
	& \stackrel{\eqref{eq:stability4}}{=} & r_br_s'(\mcd^*)\mcd^*(1-\mcd^*) - r_br_s(\mcd^*)\mcd^*  \nonumber \\
	&=& f'(\mcd^*) \underbrace{\mcd^*}_{>0}.
\end{eqnarray}
Hence, the stability of a fixed point $\mcd^*>0$ of \eqref{eq:stability} is determined by the slope of $f$ at this root. An illustration of the stability behavior is given in Fig.~\ref{fig-S1}. 

In the repulsive case ($\kappa<0$), the switch function $r_s(\mcd)$ is monotonically decreasing and therefore $f$ is monotonically decreasing in $[0,1]$ as well. Since 
$$f(0) = r_b \frac{1}{2} (1-\underbrace{\tanh(\kappa\theta)}_{\in (-1,0)} )-r_d >0,$$ 
and $f(1)=-r_d<0$, function $f$ has exactly one zero $\mcd_2^*$ and  $f'(\mcd_2^*)<0$, see Fig.~\ref{fig-S1}(a). Hence, in the repulsive case, there exists exactly one fixed point $\mcd_2^*>0$ which is stable.

In the attractive case ($\kappa>0$), the switch function is monotonically increasing and the per capita growth rate $f$ has a local maximum in $(0,1)$, as illustrated in Fig.~\ref{fig-S1}(b). If $r_s(\mcd)(1-\mcd)<r_d/r_b$ for all $\mcd$, which is the case for very large $\theta$ values, function $f$ has no zero. If values of $\mcd$ exists for which $r_s(\mcd)(1-\mcd)>r_d/r_b$, then the extistence of zeros depends on the value of $f(0)$,
\begin{equation}\label{eq:stabilityCases2}
			f(0) = \frac{r_b}{2} (1-\underbrace{\tanh(\kappa\theta)}_{\in (0,1)} )-r_d \ 
			\begin{cases}
				>0 & \mbox{ if } 1-2\frac{r_d}{r_b}<\tanh(\kappa\theta) , \\
				<0 & \mbox{ if } 1-2\frac{r_d}{r_b}>\tanh(\kappa\theta).
			\end{cases}
\end{equation}
In the first case, the function $f$ has one zero for $\mcd_2^*\approx 1$ which is stable since $f'(\mcd_2^*)<0$ for $\mcd_2^*\approx 1$. In the second case, the function $f$ has two zeros, $\mcd_2^*\approx 0$ and $\mcd_3^*\approx 1$, which are unstable and stable, respectively, since $f'(\mcd_2^*)>0$ for $\mcd_2^*\approx 0$ and $f'(\mcd_3^*)<0$ for $\mcd_3^*\approx 1$.

Finally, we conclude that in the repulsive case ($\kappa<0$), equation \eqref{eq:stability} has one unstable ($\mcd_1^*=0$) and one stable ($\mcd_2^*>0$) steady state. In the attractive case ($\kappa>0$), three cases have to be distinguished: Equation \eqref{eq:stability} has (i) one stable fixed point ($\mcd_1^*=0$) if $r_s(\mcd)(1-\mcd)<r_d/r_b$ for all $\mcd$, (ii) one unstable ($\mcd_1^*=0$) and one stable ($\mcd_2^*\approx 1$) fixed point if $1-2\frac{r_d}{r_b}<\tanh(\kappa\theta)$), (iii) two stable ($\mcd_1^*=0$, $\mcd_3^*\approx 1$) and one stable ($\mcd_2^*\approx 0$) fixed points if $1-2\frac{r_d}{r_b}>\tanh(\kappa\theta)$. 
In case (iii), the growth term \eqref{eq:stability2} shows a bistable behavior and population extinction can be observed for small densities if
\begin{equation}\label{eq:bistable}
	r_s(\mcd)<\frac{r_d}{r_b}
\end{equation}
holds.


\section*{Figures}

%
%